# Optical modulation in Ge-rich SiGe waveguides in the mid-IR wavelength range up to 11 µm

Miguel Montesinos-Ballester[1*], Vladyslav Vakarin[1,4], Joan Manel Ramirez[1,3], Qiankun Liu[1], Carlos Alonso-Ramos[1], Xavier Le Roux[1], Jacopo Frigerio[2], Andrea Ballabio[2], Andrea Barzaghi[2], Lucas Deniel[1], David Bouville[1], Laurent Vivien[1], Giovanni Isella[2], and Delphine Marris-Morini[1]

[1]Centre de Nanosciences et de Nanotechnologies, CNRS, Univ. Paris-Sud, Université Paris-Saclay, 91 120 Palaiseau, France
[2]L-NESS, Dipartimento di Fisica, Politecnico di Milano, Polo di Como, Via Anzani 42, 22100 Como, Italy
[3] now in III-V Lab, 91120 Palaiseau, France
[4] now in Nexdot, 102 Avenue Gaston Roussel, 93230 Romainville, France

*e-mail: miguel.montesinos@u-psud.fr

## Abstract

Waveguide integrated optical modulators in the mid-infrared (mid-IR) wavelength range are of significant interest for molecular spectroscopy in one hand, as on-chip synchronous detection can improve the performance of detection systems, and for free-space communications on the other hand, where optical modulators working in the atmospheric transparency windows are crucially missing. Here we report for the first time the demonstration of optical modulation in mid-IR photonic circuit reaching wavelengths larger than 8 µm. Moreover, optical modulation in an unprecedented wavelength range, from 5.5 to 11 µm wavelength is shown, relying on a broadband Ge-rich graded-SiGe platform. This first demonstration is used as a proof of concept, to experimentally confirm the free-carrier absorption effect modeling. These results pave the way towards efficient high-performance electrically-driven integrated optical modulators in the mid-IR wavelength range.

## Introduction

The mid-infrared (mid-IR) wavelength range, spanning from 2 to 20 µm, is of a tremendous interest as it covers the molecular fingerprint of many molecules, such as alkanes or greenhouse gasses, and thereby it can be used to unambiguously detect the molecular composition of a broad variety of gases, liquids or solids in a non-intrusive way. Moreover, the mid-IR range hosts both atmospheric transparency windows, at wavelengths between 3 and 5 µm and between 8 and 13 µm. Photonics Integrated Circuits (PICs) operating at those wavelengths are thus expected to have a major impact in a broad variety of high impact applications [1], such as environmental monitoring [2], astronomy [3], hazard detection [4], industrial processes control [5], medical diagnostics [6,7], thermal imaging [8] or free-space telecommunications [9].

Among the different solutions explored to develop an integrated platform at these wavelengths, silicon (Si) photonics can have a major impact by leveraging the reliable and high-volume fabrication technologies already developed for microelectronic integrated circuits. Interestingly, germanium (Ge), already used in Si photonics, is a prime candidate to extend the operating wavelength of Group IV-based PICs beyond 8 µm, potentially up to 15 µm [1, 10-12]. A whole set of passive building blocks have already been developed using different Ge and SiGe-based platforms, such as wavelength multiplexers, fiber couplers, resonators or polarization rotators [13-21]. As a next step for the development of mid-IR PICs, active devices must be achieved. More specifically, the development of an optical modulator on these Ge or SiGe-based platforms would open exciting perspectives in the context of free-space communications, since it could pave the road for a plethora of applications from short distance access network to satellite communications. Moreover, spectroscopic detection systems would also benefit from the development of an integrated mid-IR modulator, as detection sensitivity can be greatly enhanced via synchronous detection.

A large variety of high-speed optical modulators have been demonstrated using free carrier plasma dispersion effect in the near-IR wavelength range [22]. The extension of this effect to the mid-IR has been theoretically evaluated first in Si [23] and then in Ge [24]. Interestingly, the plasma dispersion effect was predicted to become more efficient at longer wavelengths. Electro-optic modulation has thus been experimentally demonstrated using diodes embedded in silicon-on-insulator (SOI) or germanium-on-silicon (GeOSi) waveguides [25-28]. Most of these demonstrations have been reported in the 2-3.8 µm wavelength range, while the development of mid-IR modulators at longer wavelength is still at its infancy with only one recent preliminary demonstration of electro-absorption modulation at 8 µm [29]. All-optical modulation has also been demonstrated based on free-carrier absorption in the 2 to 3.2 µm wavelength range [30].

The development of high performance integrated optical modulator in the mid-IR range reaching the fingerprint region beyond 6 µm and the long wave infrared (LWIR) regime beyond 8 µm still requires a careful evaluation and analysis of the electro-optic effects in SiGe materials. In that sense, experimental demonstrations appear now to be essential to confirm current models for the design of electrically-driven mid-IR optical modulators.

In this work we report the first demonstration of optical modulation in mid-IR PICs reaching wavelengths larger than 8 µm. Optical modulation in an unprecedented wavelength range, from 5.5 to 11 µm wavelength is shown, relying on a broadband Ge-rich graded-SiGe platform previously reported in [31]. This first demonstration reaching the LWIR regime is used to experimentally confirm the effectiveness of free-carrier absorption effect [24]. Furthermore, these results pave the way towards the design of efficient high-performance electrically-driven integrated optical modulators in the mid-IR range.

## Results

**Sample description.** This work relies on Ge-rich graded SiGe waveguides, similar to the ones reported in [31]. As seen in Fig. 1, the layer stack is formed by an 11 µm-thick graded layer, in which the Ge fraction increases linearly from pure Silicon (Si) up to $Si_{0.21}Ge_{0.79}$, followed by a 2 µm thick $Si_{0.2}Ge_{0.8}$ constant composition layer. This structure allows a smooth transition from pure Si to Ge-rich material while minimizing the threading dislocation density due to lattice mismatch. Since the refractive index increases linearly with the Ge content in the SiGe alloy, the mode is confined in the upper part of the epitaxial layer. As an illustration of the mode confinement effect, the normalized field intensity profile is plotted as a function of the vertical axis (z-axis) for different wavelengths between 1.33 and 11 μm (right side of Fig. 1). Such ultra-wideband light confinement is a unique property directly related to the refractive index gradient effect.

1.5 cm-long rib waveguides have been fabricated as described in the Methods section. To efficiently couple the light from the free-space optical set-up to the PIC, 50 μm-wide and 1mm-long waveguides are used as the input/output of the circuit, followed by 2 mm-long linear transitions to 7 μm-wide waveguides.

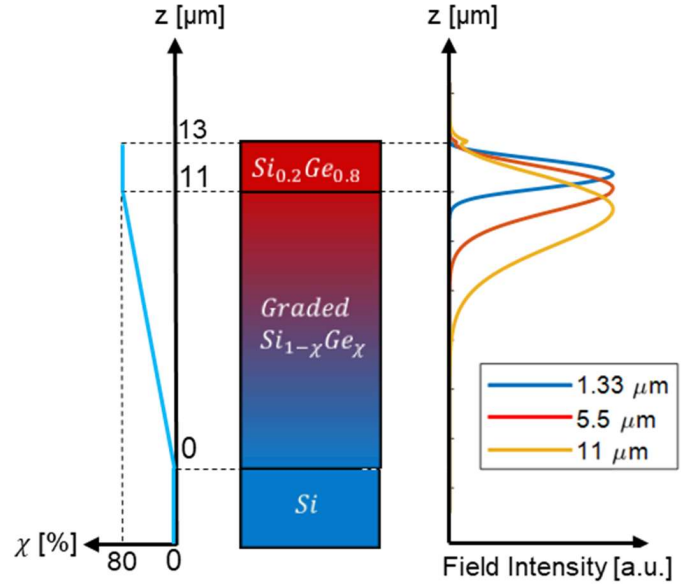

**Fig. 1** Cut view of the epitaxial layer: on top of the Si wafer the epitaxial stack is formed by a 11 μm-thick graded layer, in which the Ge fraction increases linearly from pure Si up to $Si_{0.21}Ge_{0.79}$, followed by a 2 μm thick $Si_{0.2}Ge_{0.8}$ constant composition layer. The profile of Ge fraction χ along z-axis is reported on the left-part. The normalized field profile in z-axis is reported on the right-part, for 1.33, 5.5 and 11 μm wavelengths, illustrating wideband mode confinement in the upper part of the epitaxial layer.

**Experimental evidence of optical modulation.** The goal of the experiment is to observe the effect of free carrier concentration variations on the optical absorption of Ge-rich SiGe waveguides in the mid-IR wavelength range. To generate free carriers, the waveguides are optically pumped with a near-infrared (NIR) laser at 1.33 μm wavelength. The pump wavelength is selected so that its photon energy is above the indirect bandgap energy of $Si_{0.2}Ge_{0.8}$ where most of the NIR beam is confined. Light is thus absorbed by interband absorption, generating free carriers along the waveguide. A tunable mid-IR laser beam is coupled simultaneously in the waveguide, using a free-space beam splitter to combine both beams. The effect of free carrier absorption can thus be evaluated in an all-optical modulation configuration. The scheme of the experimental setup is shown in Fig. 2 and is further detailed in Methods section. The power of the NIR beam is swept to achieve different densities of free carrier in the waveguide.

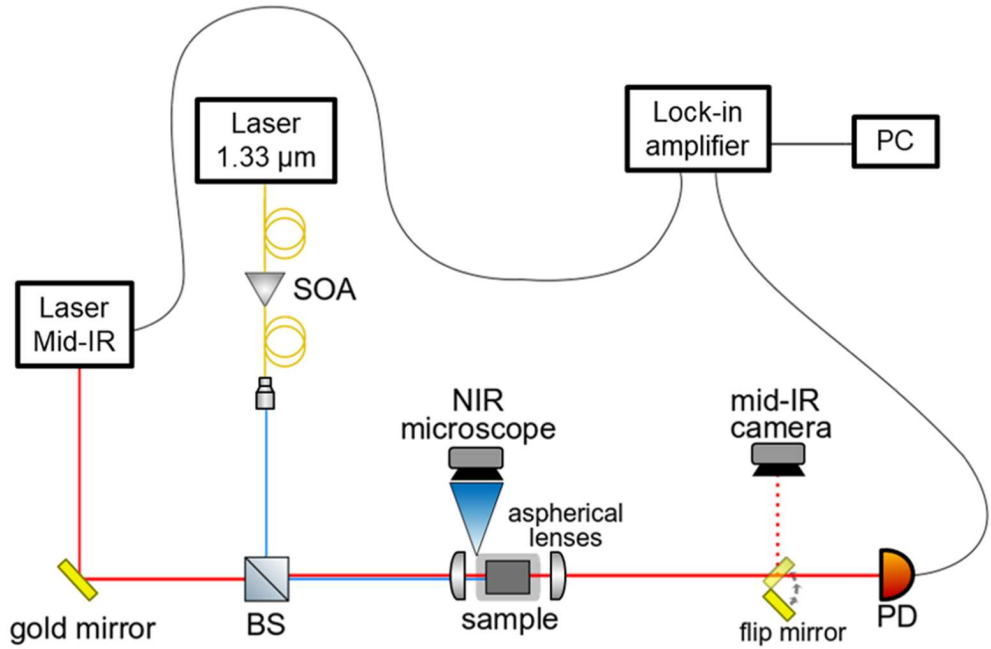

**Fig. 2** Characterization set-up scheme. A NIR beam at 1.33 µm is amplified by a Semiconductor Optical Amplifier (SOA), coupled to free-space and superposed with a tunable mid-IR beam thanks to a Beam Splitter (BS). A NIR camera is used to adjust the input coupling and the mid-IR output beam can be seen in a mid-IR camera thanks to a flipping mirror. The NIR amplitude is modulated by actuating manually a shutter in the optical path before the BS. The output mid-IR signal is collected in a Mercury Cadmium Telluride (MCT) Photodetector (PD) and a lock-in amplifier is used to improve the signal to noise ratio.

To evaluate the effect of the free carriers on the absorption of the mid-IR beam, the NIR beam is modulated by actuating manually a shutter in the optical path before the beam splitter. An example of measurement on the lock-in amplifier is reported on Fig. 3. In this example the shutter has been actuated 8 times, and the modulation of the mid-IR signal at a wavelength of 11 µm is clearly seen at each actuation, with clear on/off levels separation. A typical figure of merit for optical signals modulated in amplitude on two levels is the modulation depth (MD) that is defined in Equation 1, where $P_{off}$ and $P_{on}$ are the optical powers on the two levels. Experimentally, MD can be evaluated from $V_{off}$ and $V_{on}$ that are the mean values of the voltage measured on the lock-in amplifier for each level.

$$MD[\%] = \frac{P_{off} - P_{on}}{(P_{off} + P_{on})/2} = \frac{V_{off} - V_{on}}{(V_{off} + V_{on})/2} \qquad (1)$$

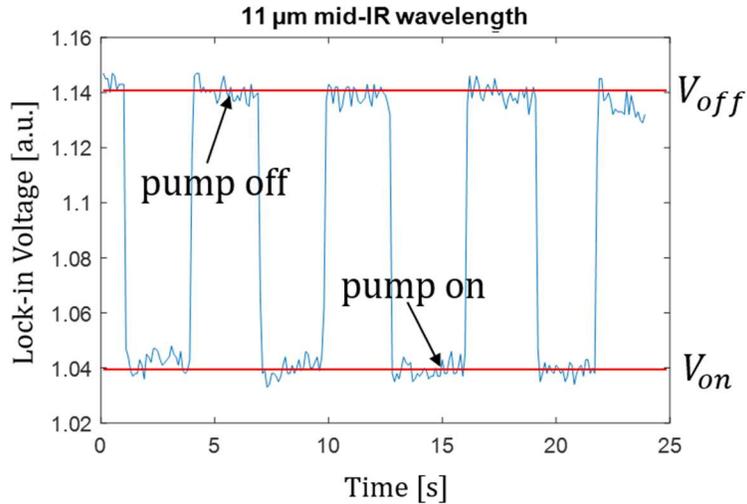

**Fig. 3** Experimental evidence of all-optical modulation at 11 μm wavelength, using a 12 mW-pump beam at 1.33 μm wavelength. The voltage at the lock-in amplifier, that is proportional to the mid-IR beam power is reported as a function of time, while the shutter is put on and off the optical path of the NIR beam. The modulation of the mid-IR signal is seen at each actuation, with clear on/off levels separation. The mean values of the measured voltages for both levels ($V_{off}$ and $V_{on}$) are deduced from this measurement and are used to calculate the MD.

A set of measurements is carried out by tuning the mid-IR laser wavelength from 5.5 to 11 μm, and for 3 values of the NIR pump powers at 1.33 μm (4, 8 and 12 mW estimated in the waveguide). The experimental results are plotted in Fig. 4. As expected, it can be seen that the MD increases when the near-IR beam power increases. Moreover, the MD first increases with the wavelength of the mid-IR signal, but then a saturation and even a slight decrease of MD is observed for wavelength values beyond 9.5 μm. While the increase of the MD with the wavelength is easily explained by the enhancement of the plasma dispersion effect as predicted by [24], the evolution of MD at longer wavelength will be further explained by the modeling of free carrier absorption in this all-optical modulation scheme.

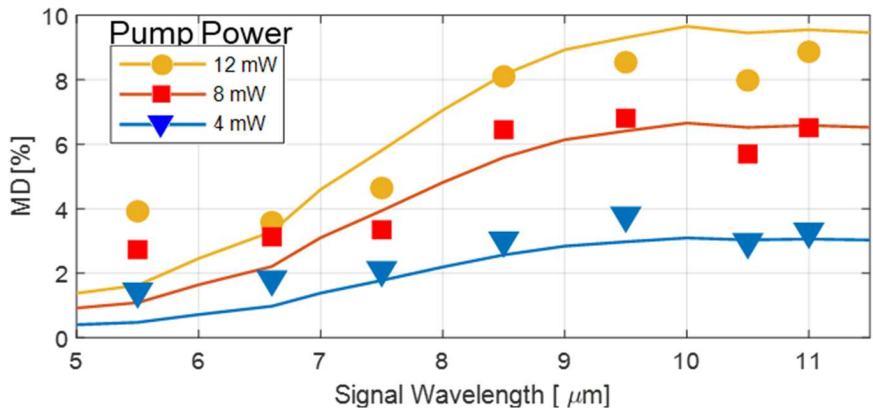

**Fig. 4** Modulation depth as a function of the signal wavelength, for different NIR pump powers (wavelength of 1.33 μm). Experimental data (dot, squares and triangles) show a good agreement with the modeling (solid lines).

**Modeling of Free carrier absorption in Ge-rich SiGe waveguides.** Once all-optical modulation reaching the LWIR regime in Ge-rich SiGe waveguides is experimentally demonstrated, it is interesting to evaluate the free carrier concentration profile in the waveguide and to compare the experimental results with the modeling of free carrier absorption effect in the mid-IR. A proper understanding of the physical effects and a good match between theory and experiment would thus be of major importance for further design of efficient electrically-driven mid-IR integrated modulator.

The modeling of the free carrier absorption effect will be detailed below. It relies on (i) the modeling of the free carrier concentration in the waveguide, (ii) the calculation of the absorption coefficient as a function of the wavelength along the waveguide (iii) the evaluation of $P_{on}$ and $P_{off}$ the power at the output of the waveguide with and without the presence of the NIR pump.

The modeling of the semiconductor carrier distribution uses the model proposed in [32]. A NIR beam at 1.33 µm wavelength propagating in (x) direction is the source of photons that generates the free carriers inside the waveguide by interband absorption. In the absence of any electrical field applied, the free carriers can undergo either free carrier recombination (described by the recombination lifetime τ) or carrier diffusion. In the (y,z) plane the free carriers distribution is described by a 2D-gaussian distribution G(y,z) with a standard deviation R (µm). If the diffusion length is larger than the NIR input beam that generates the carriers, the value of R is estimated as the carrier diffusion length $L_D = \sqrt{D\tau}$. Finally the excess carrier density Δn(x,y,z) is thus given by the following equation:

$$\frac{\Delta n(x,y,z)}{\tau} - D\frac{d^2\Delta n(x,y,z)}{dx^2} = \alpha\, \Phi_0.\, G(y,z) e^{-\alpha x} \qquad (2)$$

where τ is the free carrier recombination lifetime, D is the ambipolar diffusion coefficient [33], $\alpha$ is the NIR absorption coefficient, $\Phi_0$ is the incident photon flux, G(y,z) is the normalized gaussian distribution function describing the free carrier distribution in the (y,z) plane with standard deviation R and x is the position along the propagation axis. $\Phi_0$ can be calculated from the input power by $\phi_0 = \frac{P_{NIR}}{h\, c/\lambda_{NIR}}$, where $P_{NIR}$ is the power of the NIR beam, h is the Plank constant, c is the light speed in vacuum and $\lambda_{NIR}$ is the wavelength of the NIR beam.

Assuming a surface recombination rate $S_0$ at x=0 and a free carrier density that vanishes at x=+∞ as boundary conditions [32], the carrier density can be calculated as:

$$\Delta n(x,y,z) = \frac{\alpha\, \tau\, \Phi_0 G(y,z)}{(\alpha L_D)^2 - 1}\left(\frac{\frac{S_0}{D} + \alpha}{\frac{S_0}{D} + 1/L_D} e^{-x/L_D} - e^{-\alpha x}\right) \qquad (3)$$

Free carriers are generated by the optical absorption at 1.33 µm wavelength and most of this absorption occurs in the $Si_{0.2}Ge_{0.8}$ top constant composition layer. The absorption coefficient is thus 20 $cm^{-1}$ [34]. The ambipolar diffusion coefficient can be estimated to be 65 $cm^2 s^{-1}$ [33, 35]. In terms of recombination velocity at the input surface, we use values that have been reported for Ge surface [36], i.e. $S_0 = 250\ cm/s$. The carrier recombination time can strongly vary with growth conditions and fabrication process. We thus decided to fit this value by comparing the experimental measurements and the simulation results, both shown in Fig. 4. A value of τ = 15 ns is obtained, which is consistent with the results reported in [30], where a value of a value $\tau$ of 18 ns has been estimated for Ge-on-Si waveguides. The corresponding carrier diffusion length is thus 10 μm. This value has been used as the standard deviation R of the free-carriers distribution G(y,z), as it is larger than the NIR beam spot size.

Based on these data, carrier density distribution along the propagation axis can be calculated with Eq. 3 at any (y,z) position. The electrons and holes concentrations in the center of the normalized Gaussian distribution G(y,z) and along x-axis are reported in Fig. 5 for different values of the NIR powers. The maximal carrier concentration in the waveguide entrance is $3.9 \times 10^{15}$ cm$^{-3}$ for 12 mW input power. The carrier distribution decays rapidly along the propagation direction up to negligible values after a few millimeters of propagation. This indicates that the NIR beam is strongly absorbed as soon as it is coupled in the PIC, which implies that most of the modulation will then occur in the 50 μm-wide 1 mm-long waveguide input waveguide.

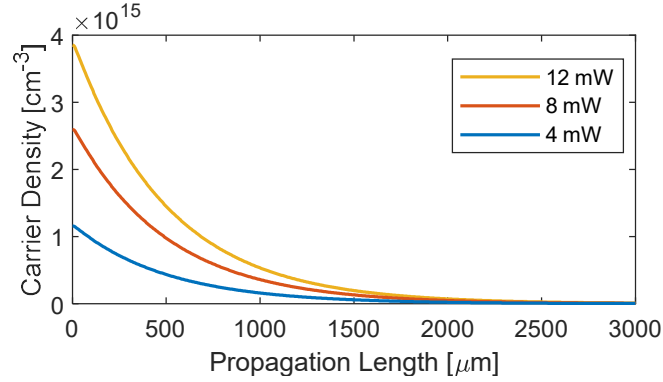

**Fig. 5** Carrier density distributions along the propagation direction (x-axis) at the center of the (y,z) gaussian distribution G(y,z) and for different pump powers coupled in the waveguide: 4, 8 and 12 mW.

Once the carrier distributions are obtained, the absorption coefficient due to free carriers can be estimated. As most of the optical mode propagates in the Ge-rich part of the epitaxial growth, the model developed in Ref. [24] is used, where the effect of free carrier on the optical absorption is evaluated for Ge. Absorption variation is thus given by the following equation were c1(λ), c2(λ), c3(λ), c4(λ) are obtained from [24].

$$\Delta\alpha(\lambda) = c_1(\lambda)\, \Delta n_e^{c_2(\lambda)} + c_3(\lambda)\, \Delta n_h^{c_4(\lambda)} \qquad (4)$$

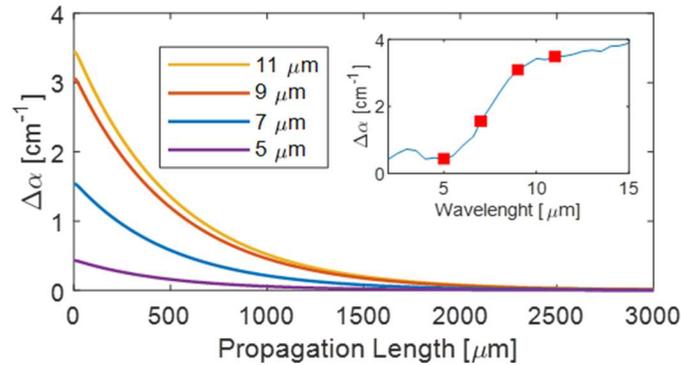

**Fig. 6** Absorption coefficient due to free carriers along the propagation direction (x-axis) at the center of the (y,z) gaussian distribution , for a pump power of 12 mW and for different mid-IR wavelengths. Inset: absorption coefficient increase as a function of the wavelength for a carrier density of $3.9 \times 10^{15}$ cm$^{-3}$, calculated by Eq. 4 and considering the parameters c1(λ), c2(λ), c3(λ), c4(λ) reported in Ref. [24] for Ge.

The absorption coefficient increase due to free carriers along the propagation direction can be deduced from Eq. 4, using the calculated carrier density distribution. The result is reported in Fig. 6 for an input pump power of 12 mW. As expected, the absorption mainly occurs in the first millimeter of propagation. The absorption coefficient increase is also reported as a function of the wavelength in the inset of Fig. 6, using a carrier concentration for electrons and holes of 3.9×10$^{15}$ cm$^{-3}$. A strong increase of the absorption coefficient is observed between 5 and 9 µm, but it suffers from a saturation for wavelengths larger than 9 µm.

To evaluate the modulation of the mid-IR signal, light confinement of the mid-IR beam must be considered. As most of the absorption occurs in the first millimeter of propagation, i.e. in the 50 µm-wide waveguide, a non-uniform gaussian distribution in z and y direction is used, considering different values for the vertical and horizontal standard deviation ($R_{mid-z}$ and $R_{mid-y}$ respectively). In the vertical direction, the mode size and position strongly depend on the wavelength (right side of Fig. 1). Numerical simulations are thus used to estimate both $R_{mid-z}(\lambda)$ and the position of the optical mode center in the vertical direction. In the lateral dimension, the confinement is much lower because of the 50 µm-wide input waveguides. Thus, the beam size is roughly given by the spot size at the output of the coupling lens. $R_{mid-y}$ is then estimated experimentally with a camera at the waveguide output, and a value of 12 µm at 5.5 µm wavelength is obtained. For larger wavelengths, the following relation is used : $R_{mid-y}[\mu m] = R_{5.5\,\mu m-y} \sqrt{\lambda_{mid}[\mu m]/5.5}$ where $\lambda_{mid}[\mu m]$ is the signal wavelength and $R_{5.5\,\mu m-y}$ takes a value of 12 µm.

It can be noted that when the mid-IR wavelength increases, the larger MIR beam and its deeper vertical displacement in the graded layer are both responsible of a reduction of the overlap between the beam and the free carrier distributions, leading to a reduction of MD.

Finally, the evolution of the power at mid-IR wavelengths along the propagation direction can be calculated by integrating the mid-IR signal distribution with the locally calculated free-carrier-based absorption coefficient. Typical propagation loss of 2.5 dB/cm [31] is also considered all-along the waveguide and added to the free-carrier-based absorption. As an example, the evolution of the power inside the waveguide is reported in Fig. 7 at 11 µm wavelength. Without NIR pump, the mid-IR power experiences a long exponential decay due to the constant propagation losses. With 12 mW pump at 1.33 µm, the effect of the free carriers can be seen at the waveguide entrance by a stronger power decrease in the first millimeter. Then the slope of the transmission as a function of the length becomes equal to the case without free carriers, meaning that only the contribution from classical propagation loss can be seen. From this calculation the MD can be obtained by evaluating $P_{on}$ and $P_{off}$ at the waveguide output.

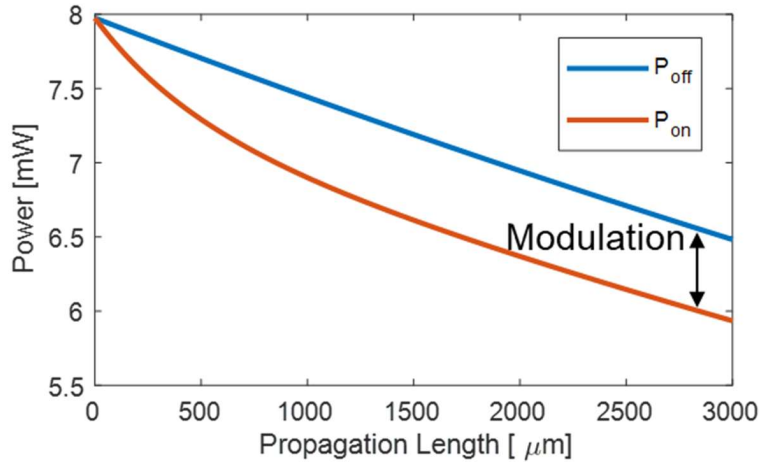

**Fig. 7** Evolution of the power of the signal at 11µm wavelength, along the propagation direction, when a 12 mW NIR beam at 1.33 µm is applied (blue) or not (orange). 2.5 dB/cm propagation losses are considered in both cases. The effect of the free carriers can be seen in the first millimeter, by an increase of the losses. The modulation can be deduced at the waveguide output.

This method has been applied for different mid-IR wavelength from 5.5 to 11 µm and for different NIR pump power (4, 8 and 12 mW). The results are plotted as solid lines in Fig. 4 together with the experimental measurements in dots.

A good agreement between experiments and simulations can be observed. The modeling reproduces the evolution of MD with the signal wavelength, i.e. a strong increase of MD with the wavelength from 5 to 9.5 µm, followed by a saturation and even a decrease of MD beyond 9.5 µm. This evolution is explained by a combination of both the saturation of the plasma dispersion effect modeled in [24] and reported in the inset of Fig. 6 and a reduction of the overlap between carriers and the mid-IR beam when the wavelength increases, as previously discussed. This experiment is the first experimental demonstration of free-carrier absorption effect over such a wide mid-IR spectral range.

## Discussion

This work reports the first experimental demonstration of optical modulation in a mid-IR PIC carried out in LWIR regime, up to 11 µm. The reported modulation performances are modest in term of MD, but this is easily explained by the decrease of free carrier concentration inside the waveguide along the propagation. Interestingly, experimental results are in a good alignment with free-carrier electro-absorption predictions [24], which is a first step towards the realization of efficient high performance electrically-driven integrated optical modulator in the mid-IR range. Indeed, the use of a PIN or PN diode embedded in the waveguide to inject free carriers all-along the modulator active region can be envisaged. Free carrier concentration variations in the range of $10^{16}$ cm$^{-3}$ can then be expected. Absorption variation around $\Delta\alpha = 6$ cm$^{-1}$ can then be predicted at 8 µm wavelength, allowing to expect more than 5 dB extinction ratio for a 2 mm-long modulator.

These result paves the way towards the design of efficient mid-IR modulators and confirm the huge potential of Ge-based photonic circuits as a promising platform for the integration of mid-IR photonics functions, as it could benefit from low loss waveguides, compact photonics structures, together with the integration of active devices such as optical modulator. The realization of complete functional systems in the MIR and LWIR paves the way for a plethora of applications, from mid-IR spectroscopy, sensing, to free-space telecommunications.

## Methods

**Fabrication.** The epitaxial growth is performed on a silicon wafer by low energy plasma enhanced chemical vapor deposition (LEPECVD) at a rate of 5-10 nm/s. Then, a set of straight waveguides are fabricated by laser lithography and reactive-ion inductively coupled plasma etching (RIE-ICP). The typical waveguide width is 7 µm and the etching depth is 3.8 µm. These dimensions have been measured with a scanning electron microscope (SEM). To efficiently couple the light from the free-space optical set-up to the PIC, 50 µm-wide and 1 mm-long waveguides are used as the input/output of the PIC, followed by 2 mm long linear transition to the 7 µm-wide waveguide.

**Measurements setup.** The scheme of the experimental setup is shown in Fig. 2. We use a pulsed quantum-cascade tunable laser from 5.5 up to 11 µm wavelength range, with a repetition rate of 100 kHz and duty cycle of 5%. The mean power varies with the wavelength, with a maximum value of 30 mW at 6.25 µm. The mid-IR beam is spatially superposed with a continuous wave (CW) 1.33 µm wavelength laser beam, thanks to a beam splitter (BS). The NIR laser is first amplified by a Semiconductor Optical Amplifier (SOA) and then sent to free-space by a collimator lens. Both beams are coupled together in and out of the PIC thanks to broadband ZnSe aspheric lenses. The NIR power is swept by variation of the SOA current to achieve different free carrier concentrations in the waveguide. It is evaluated by measuring the power of the NIR beam with a broadband power-meter before coupling to the sample and estimating 5 dB of coupling loss. The measurement is reported for 4, 8 and 12 mW coupled on-chip, evaluated the mid-IR modulated signal is collected from the waveguide output and sent to a Mercury Cadmium Telluride (MCT) detector. A lock-in amplifier triggered by the pulsed mid-IR laser is used for signal-to-noise ratio improvement. The output voltage of the lock-in amplifier is thus proportional to the power of the collected mid-IR beam. A mid-IR camera is also placed thanks to a flipping mirror to ensure a correct alignment and coupling.

## Acknowledgements

This work was supported by the European Research Council (ERC) under the European Union's Horizon 2020 (No 639107 - INsPIRE). The fabrication of the device has been performed in the Platforme de Micro-Nano-Technologie/C2N, which is partially funded by the "Conseil Général de l'Essonne". This work was partly supported by the french RENATECH network.

A. Ballabio and A. Barzaghi acknowledge financial support by the TEINVEIN project funded by POR FESR 2014-2020 (ID: 242092). J. Frigerio acknowledges financial support by the European Union Horizon 2020 FET project microSPIRE (ID: 766955).

## Author contributions statement

J. M. R. proposed the experiment. M.M.B. and D.M.M developed the theoretical model and data analysis. J.F., A.B., A.B. and G.I. carried out the epitaxial growth. M.M.B., V.V., D. B., Q. L. and X.L.R. fabricated the structures. M.M.B. and L. D. performed the measurements. M.M.B., C.A.R., G. I., L.V. and D.M.M. discussed the results and wrote the manuscript.

## Additional information

Competing Interests: The authors declare that they have no competing interests